\documentstyle[aps,prl,twocolumn]{revtex}
\begin{document}
\title{Low energy dynamics of the one dimensional multichannel Kondo-Heisenberg Lattice}
\author{N. Andrei and E. Orignac}

\address{ Serin Laboratory, Rutgers University, P.O. Box 849,
Piscataway, NJ 08855-0849, USA} 
\wideabs{
\date{\today}
\maketitle

\begin{abstract}
We determine {\it exactly} the fixed point Hamiltonian of the 
one dimensional multichannel Kondo-Heisenberg lattice model
 for any number of channels $N \ge 2$. An anomalous singlet with non trivial 
internal dynamics is generated.
We compute the correlation functions of the various conventional and
 unconventional order parameters of the system and find that for $ N\le 4 $
the composite order parameter induce the dominant instabilities.
\end{abstract}
}
\vspace*{0.2truecm}

The Kondo lattice is one of the most challenging problems in
contemporary theoretical condensed matter physics. If the single impurity
Kondo problem, a local moment antiferromagnetically coupled
to conduction electrons, is by now well understood thanks to a variety
of theoretical
techniques\cite{kondo_papers}
the problem of a regular three-dimensional
 array of local moments in a metal (the Kondo lattice) still defies
theoretical analysis. The difficulty stems from the fact that there
are two competing effects, the tendency of the local moments to form
singlets with the conduction electrons (or more complicated
states\cite{nozieres_multichannel} if 
more than one band is available) and the tendency of the moments to
order due to the RKKY interaction mediated by the conduction
electrons. Besides its intrinsic theoretical interest for the general
theory of strongly correlated fermions, the Kondo lattice model is
believed  to capture the non-Fermi liquid physics of a class
of rare-earth or actinide compounds\cite{maple_nfl_review}
 in which f-shell local moments 
couple to the electron bands. In this context\cite{cox_kondo_review}, 
it has been suggested that two-channel
Kondo model could be relevant in Ce$^{3+}$ or U$^{4+}$ based alloys such
as UBe$_{13}$ and CeCu$_2$Si$_2$. This hypothesis is supported by
experiments on diluted alloys that show a two-channel single impurity
behavior in the specific heat or magnetic susceptibility. A full
understanding of the multichannel Kondo lattice would also be of
great significance for experiment.

Many problems of
strongly correlated fermions can be treated very effectively in one dimension
thanks to
powerful analytical~\cite{schulz_houches_revue} and numerical~\cite{white} techniques available there.   
In particular, the full phase diagram of the one-dimensional single
channel Kondo Lattice has been determined\cite{tsunetsugu_kondo_1d}.
An insulating state with a spin gap
obtains at half filling. Away from half filling a small
Kondo coupling  produces a paramagnetic metal and a large one leads to
ferromagnetism. Adding an antiferromagnetic Heisenberg interaction between the local moments
(Kondo-Heisenberg lattice) produces  a spin-gapped metal
\cite{sikkema_spingap_kondo_1d} with
non-conventional 
pairing fluctuations\cite{exotic_pairing}.

In this Letter, we investigate the ground state, excitations and dominant 
instabilities of the multichannel one dimensional Kondo-Heisenberg 
lattice model for incommensurate
fillings and zero temperature. 
We find that  screening of the local moments occurs via the 
 "chiral stabilization" mechanism \cite{andrei_chiral_nfl}, resulting in a
 ground state that is  a chiral non-Fermi
liquid  we call \emph{coset singlet}.
We show that the dominant instability  for a low number of channels
$N \le 4$  is of the
non-conventional pairing type as in the single channel
case\cite{exotic_pairing} and that  for a larger number of channels, one
recovers conventional pairing instabilities as the dominant ones. 
 Fermi Liquid physics is regained in the limit of very large number of
 channels. We also discuss the
effect of channel anisotropy.

The Hamiltonian of the one dimensional multichannel Kondo-Heisenberg lattice  is:
\begin{eqnarray}\label{eq:lattice_hamiltonian}
H&=&-t \sum_{i,n,\sigma}(c^\dagger_{i,n,\sigma} c_{i+1,n,\sigma} +
c^\dagger_{i+1,n,\sigma} c_{i,n,\sigma}) \nonumber\\&+& \lambda_H \sum_i
\vec{S}_i\vec{S}_{i+1} + \lambda_K   \sum_{i,n} \vec{S}_i . c^\dagger_{i,n,\alpha}\vec{\sigma}_{\alpha,\beta}c_{i,n,\beta},
\end{eqnarray}
where $\sigma=\uparrow,\downarrow$ is the spin of an electron, $n=1
,\ldots,N$ is the channel index and $\vec{S}_i$ is a localized spin
$1/2$. $t$ is the bandwidth, $\lambda_H>0$ the Heisenberg coupling of
the localized spins and $\lambda_K>0$ the Kondo coupling. 
  
To study the low energy physics of the model we follow the standard strategy\cite{affleck_tsvelik}:  keep only linear electron modes around $\pm k_F$  captured
in terms of the left and right moving fields  $\psi_{L,R,\sigma,n}(x)$,
as well as the low-lying spin-lattice modes around $\frac{\pi}{a}$.
The low energy physics is then described by a continuum
Hamiltonian  expressed using non-abelian
bosonization\cite{witten_wz,affleck_tsvelik} as a quadratic 
form of various currents: the  $SU(2)_1$ 
(left and right) spin currents $\vec{\sigma}_{L,R}$ of the local moments, the $SU(2)_N$
(left and right) spin current $ \vec{S}_{L,R}(x)$ of the electrons and the electron charge and
channel currents which decouple from the spin currents in the continuum. The
notation $SU(M)_k$ indicates that the currents (generically denoted $J_{L,R}^a$) satisfy
 a Kac-Moody algebra: 
$[J^a_L(x),J^b_L(x')]=\delta(x-x') f^{abc} J^c_L(x) + \frac{k}{2\pi}
\delta'(x-x')$, where $f_{abc}$ are the structure constants of  the $SU(M)$  
Lie algebra. A similar 
relation holds for the currents $J_R$, while the $J_R$ and
 $J_L$ currents commute. 
The KM algebra generates a
 conformal field theory
 (CFT) known in Lagrangian form as the
Wess-Zumino-Novikov-Witten\cite{witten_wz} 
 (WZNW) model. All the physical 
 operators of the original lattice theory
(\ref{eq:lattice_hamiltonian}) can be 
 expressed in terms of the primary or descendant fields of this
 CFT allowing the calculation of the
asymptotic behavior of their correlators. This is at the heart of all
 applications of non-abelian bosonization to condensed matter physics.

Consider the Hamiltonian with $\lambda_K=0$. The electrons are
free, and in the basis described
above  charge, spin and channel excitations propagate independently of each other,
and decouple from the local-moment excitations. 
Turning on a small antiferromagnetic
 Kondo coupling $0< \lambda_K \ll t, \lambda_H$ at
 incommensurate filling,  the charge and flavor
excitations remain  decoupled from the spin excitations. 
We can thus restrict ourselves to the latter,
described by the Hamiltonian:
\begin{eqnarray}\label{eq:continuum_hamiltonian}
H&=&\int \left[\frac{2\pi v}{N+2}  (\vec{S}_L.\vec{S}_L +
\vec{S}_R.\vec{S}_R)+ \frac{2\pi v_s}{3} (\vec{\sigma}_L.\vec{\sigma}_L
+\vec{\sigma}_R.\vec{\sigma}_R) \nonumber \right. \\
&+&\left. \lambda^f_K   (\vec{S}_L.\vec{\sigma}_L + \vec{S}_R.\vec{\sigma}_R) +
\lambda^b_K   (\vec{S}_L.\vec{\sigma}_R + 
\vec{S}_R.\vec{\sigma}_L)\right]
\end{eqnarray}
where we dropped irrelevant terms such as $\vec{\sigma}_L.\vec{\sigma}_R$
The coupling $\lambda^f_K$ 
describes the forward scattering, $\lambda^b_K$ the backward 
scattering and $\lambda^f_K=\lambda^b_K=\lambda_K$ to begin
with. Under RG transformations $\lambda^f_K$ does 
not flow near the weak coupling fixed point and merely renormalizes
 the velocity. On the other hand $\lambda^b_K$ is relevant and
 drives the system to a new strong coupling fixed point.
We  will thus take
$\lambda^f_K=0$ and $v_s=v$ in (\ref{eq:continuum_hamiltonian})
 and determine the fixed point.
We will later  prove
that the neglected terms  are  irrelevant near
 the strong coupling fixed point.

Under these circumstances the spin sector 
is  described by $H = H_1 + H_2 $ where:
\begin{eqnarray}\label{eq:chiral_theory}
H_1& =& \int dx \left[ \frac{2\pi v}{N+2} \vec{S}_R \vec{S}_R + \frac{2\pi
v}{3} \vec{\sigma}_L \vec{\sigma}_L + \lambda_K
\vec{S}_R. \vec{\sigma}_L\right] \nonumber \\
H_2 &=& \int dx \left[ \frac{2\pi v}{N+2} \vec{S}_L \vec{S}_L + \frac{2\pi
v}{3} \vec{\sigma}_R \vec{\sigma}_R + \lambda_K \vec{S}_L. \vec{\sigma}_R\right].\nonumber
\end{eqnarray}
In $H_1$ (resp. $H_2$), the left (resp. right) branch of the $SU(2)_N$
WZNW model is coupled to the right (resp. left) branch of the
$SU(2)_1$ WZNW model, leading to a chiral asymmetry of $H_1$ (resp. $H_2$).
We readily identify the strong coupling fixed point of
(\ref{eq:chiral_theory}) exactly  
via ``chiral stabilization''
\cite{andrei_chiral_nfl}.
The chiral asymmetry in $H_1$ or $H_2$ is
 invariant under the RG flow and 
characterizes the fixed point. We find this way that the  
theory  (\ref{eq:chiral_theory}) flows under
RG to a fixed point theory that is
the product of a coset
theory\cite{goddard_unitar_repres_viras}
 by a WZNW theory:

\begin{equation}\label{eq:fixed_point_theory}
H^* = 
\frac{SU(2)_1\times SU(2)_{N-1}}{SU(2)_N} \otimes   SU(2)_{N-1}
\end{equation}
where the coset theory, $\frac{SU(2)_1\times
SU(2)_{N-1}}{SU(2)_N}$  describes 
the local moment spin
sector, and  the $SU(2)_{N-1}$ WZNW theory describes the electron spin
sector. Note that at the fixed point the left and right components is $H_1$
and $H_2$ are recombined  and chiral symmetry is globally
preserved.
 
What is the physics around the fixed point? The coset  theory describes
a spin singlet which
the local moments 
form with the electrons. It is 
a new type of a singlet, a {\it coset singlet}:
a fraction  $\frac{6}{(N+1)(N+2)}$ of the local moment  
``modes'' are paired
with  the same number of electron  spin  ``modes''. Thus 
the system loses twice this amount of degrees of freedom
as seen in  the total specific heat (including channel and charge
degrees of freedom):
\begin{eqnarray}
C_{\text{total}}=\frac{\pi}{6} \left(2N+1-\frac{12}{(N+1)(N+2)}\right) T.
\end{eqnarray}
The susceptibility is given by,
\begin{eqnarray}
\chi=\frac {1} {2 \pi v} (N-1)
\end{eqnarray}
and the Wilson Ratio: $R_w=\frac{2N+1 -\frac{12}{(N+1)(N+2)}}{N-1}$.
The coset singlet  still retains  degrees of freedom whose number is given by
the central  charge of the theory $c=1-\frac{6}{(N+1)(N+2)}$. 
This fraction decreases with the number of
 channels $N$ since  it is ``easier'' to 
form the singlet when $N$ increases. This also shows up in the 
effective coupling of the electrons 
to the local moments which decreases with
with the number of channels, $\lambda^*_K \sim 1/N$.

Consider the two-channel case, \cite{realiz} .
For $N=2$, local moments  are
described by a   $\frac{SU(2)_1\times
SU(2)_{1}}{SU(2)_2} =\text{Ising}$ theory, or equivalently
 by a  Majorana fermion. 
Such a Majorana fermion picture is very appealing
since it is well known \cite{kondo_2channel} that the
the single impurity two channel Kondo model be described at the fixed point
by a local
Majorana fermion degree of freedom. 
These Majorana fermions form a band when coupled
with each other, thus suppressing the single impurity 
residual entropy at $T=0$.

Having obtained the low energy theory (\ref{eq:fixed_point_theory})
 describing the spin sector, we proceed to express the original operators 
 in terms of the operators of the
fixed point theory. This will enable us to check that the operators
 we  discarded in (\ref{eq:continuum_hamiltonian}) are indeed
irrelevant at the strong coupling fixed point as well as
 determine the dominant instability of the Multichannel
Kondo-Heisenberg Lattice. The needed identification of
operators as well as the calculation of the scaling dimensions was
done in Ref. \onlinecite{andrei_chiral_nfl}.
Let us summarize briefly the method. To obtain the conformal weight of
 a given operator, we first decompose it into a product of operators
 belonging to each of the two decoupled chiral theories. 
Then, for each chiral theory, we  decompose operators of
 $SU(2)_{N} \otimes SU(2)_{1}$ on operators of $SU(2)_{N-1}$ in an
 expansion formally similar to the Clebsch-Gordan expansion, the
 role of the Clebsch-Gordan coefficients being played by operators of
 the coset theory\cite{andrei_chiral_nfl}. 
The operator with the lowest scaling dimension in this expansion
 is then retained as the fixed point form of the original operator. 

 The
results are summarized in the table \ref{tab:conformal_weights} for
the theory described by $H_1$. The conformal weights of the theory
described by $H_2$ are obtained by interchanging $L$ and $R$. 
These  conformal weights are such that the operators
 $\vec{S}_L(x).\vec{\sigma}_L(x)$,
  $\vec{S}_R(x).\vec{\sigma}_R(x)$ and $(v_s-v)
( \vec{\sigma}_R. \vec{\sigma}_R +\vec{\sigma}_L. \vec{\sigma}_L)$
that we have previously discarded are indeed irrelevant (marginally irrelevant for $N=2$). This proves
the self-consistency of our treatment.

The fixed point  is a non-Fermi liquid.
The Green's functions of the right moving fermions is given by:
\begin{eqnarray}
\langle T \psi_R(x,t) \psi_R^\dagger(0,0) \rangle \sim \frac 1
{(x-vt)^{1+\frac{1}{2}\delta_N} (x+vt)^{\frac{1}{2}\delta_N}} \nonumber
\end{eqnarray}
where $\delta_N =\frac{3}{ (N+1)(N+2)} $. Here we have taken, for simplicity,
charge, spin, and channel velocities to be equal, and combined exponents from all sectors, neglecting the non-universal Luttinger interaction in the charge sector (it can be easily taken into account \cite{andrei_chiral_nfl}). Further, if a
 contribution in the  channel sector is generated it is ferromagnetic and
flows to zero. $G_L(x,t)$ is obtained by
 replacing $x\pm vt$ by $ x\mp vt$. We note that  a weak singularity appears
 at the Fermi level $k_F$, and there is no large Fermi surface. Also note that all dimensions tend to their Fermi liquid values  in the limit of large number of channels.

We now examine  the possible order parameters of the system.
Beginning with the localized moments, we have $\vec{S}_i=
\vec{\sigma}_L(x)+\vec{\sigma}_R(x) + e^{\imath \pi x/a} \vec{n}(x)$,
where $\vec{n}(x)$, the staggered magnetization, is given by $
\vec{n}(x)=\frac 1 2 \sum_{\alpha,\beta} \tilde{g}^\dagger_{R,\alpha}
\vec{\sigma}_{\alpha,\beta} \tilde{g}_{L,\beta}$, where 
$\tilde{g}_{R,\alpha}, \;(\tilde{g}_{L,\alpha})$ are  the  right (left) WZNW fields. We find, 
\begin{equation}
\langle \vec{n}(x). \vec{n}^\dagger(x')\rangle \sim \frac 1
{ |x-x'|^{1+ \frac{6}{N+1}} }
\end{equation}

More order parameters are available in the electron sector.
 The order parameters for charge density wave (CDW),
spin density wave (SDW), singlet (SS) and triplet (TS) superconducting
 are defined in the Table and their dimension is
 given, from which it follows that they are degenerate and fall with a power
 $2+2\delta_N$. As the 
 fluctuations of these order parameters are   
weaker than in the one dimensional metal we are lead
 to investigate the possibility of dominant fluctuations
associated with a non-conventional order
parameter\cite{exotic_pairing}  odd-frequency singlet pairing\cite{odd_freq_pairing}
(c-SP) and composite charge-density wave order (c-CDW): $
O_{\text{c-SP}}=\vec{n}.\vec{O}_{TS}$ and $ O_{\text{c-CDW}}=\vec{n}.\vec{O}_{\text{SDW}}$. The composite operators have momentum $\pm \frac {\pi} a $
and $2k_F \pm \frac \pi
a$ respectively, and are  associated with the gapless excitations predicted by
Yamanaka et al.\cite{yamanaka_luttinger_thm}; their  correlations are expected
 display quasi-long range order. 
Indeed,  they decay with the power $3-\frac{6}{N+2}$.

We observe that for $N \le5$, the composite order parameters
have the most divergent correlations. In this case a
large enough fraction of electron spin degrees of freedom is bound the
the local moments to suppress the conventional  order parameters and
enhance the composite ones. This situation is  similar to
one-channel one dimensional Kondo
Lattice case\cite{exotic_pairing} where composite pairing operators
are also dominant.
For $N=5$, the two types of order are degenerate
\emph{at the fixed point}, and one needs to
describe also the approach to the fixed point starting from the bare
Hamiltonian to determine the order. 
For $N\ge 6$, the fraction of electron
spin degrees of freedom  bound to a local moment is
insufficient to permit  composite order parameters to dominate the
conventional ones, thus recovering the Fermi liquid limit. 

Thus far, we  assumed channel isotropy. In the single impurity problem
\cite{nozieres_multichannel} channel anisotropy is a relevant
perturbation, so we have to investigate its effect for a lattice. 
Assume
 $N_1$ channels couple to the local moments with coupling
strength $\lambda_K^1$ and $N_2=N-N_1$ channels couple with strength
$\lambda_K^2$. Then, the spin excitations of the $N_1$ channels are
described by a $SU(2)_{N_1}$ KM algebra, whereas those of the $N_2$
remaining channels are described by a $SU(2)_{N_2}$. 
For $\lambda_{K}^1 \ll \lambda_{K}^2$, {\it coset} screening  occurs between
the $N_1$ channels and the local moments. The resulting theory is:$
\frac{SU(2)_1\otimes SU(2)_{N_1-1}}{SU(2)_{N_1}} \otimes SU(2)_{N_1-1}
\times SU(2)_{N_2}$, with  central charge: $
c=2N+1-\frac{12}{(N_1+1)(N_1+2)}$ smaller than $c= 2N+1-\frac{12}{(N+1)(N+2)}$
the charge of the symmetric fixed point. It is  the stable
  fixed point having the lower number of degrees of freedom\cite{zamolodchikov_c_thm}. This implies
 that channel anisotropy is always relevant as in the single
impurity problem. 

\smallskip

We  have derived the fixed point theory describing the one
dimensional multichannel Kondo-Heisenberg lattice. This fixed point
 is in the class of chiral non-fermi
liquids. We showed that the dominant
instability is of the composite
pairing type for a number of channels smaller than five and of the
conventional pairing type otherwise.
Our results were obtained in the limit of weak Kondo
coupling, at zero temperature and away
from half-filling. At half-filling, it is known that a spin gap
develops only for $N-2S$ integer\cite{tsvelik_mck_comm}.

There are many directions for future research:
At finite temperature, the  behavior  we 
described must disappear above the larger of the Kondo 
Temperatures $T^e_K$ and
$T^s_K$ characterizing the electron and local moment sectors, respectively. 
Also the problem of exhaustion\cite{nozieres_exhaustion}, how a large
 concentration of
impurities reduces the Kondo scale, is still unresolved. The conformal
field theory approach of our paper is not able to settle this issue. 
However, the model (\ref{eq:continuum_hamiltonian}) happens to be Bethe Ansatz
integrable. This will allow  us to settle
the issue of the Kondo scales and their dependence
on the filling -  exhaustion - and discuss the full crossover to the zero
temperature behavior. A simple generalization is to replace the spin-1/2
chain by an integrable spin-$S$ chain of local moments, which would
 lead to a replacement of the $SU(2)_1$ by $SU(2)_{2S}$ Kac-Moody algebra.
 Another line of future research
relates to coupling two or more Kondo chains to investigate the effects
of the unstable fluctuations.
A last question is whether the picture we have obtained 
 persists for $\lambda_K \gg
t,\lambda_H$ or very low density and if not what is the strong $\lambda_K$
regime. Presumably, such regime would be a ferromagnetic Nagaoka--like
state as in the single channel case. 
Besides the various generalizations of the one dimensional
 Kondo-Heisenberg lattice problem, it would be interesting to
determine whether the physics of the Kondo-Heisenberg problem persists
in the Kondo limit $\lambda_H=0$. It is known that this is not so 
in the one-channel 
case\cite{tsunetsugu_kondo_1d,sikkema_spingap_kondo_1d}. However, since in the multichannel
case there is only a partial screening of the electrons by the spins,
one may expect that a RKKY interaction could be generated even in the
pure Kondo problem putting it in the universality class of the
Kondo-Heisenberg problem. 

We are grateful
to P. Lecheminant, P. Azaria, O. Parcollet and  A. Rosch for
illuminating discussions and comments. 
E. O. acknowledges support from NSF under Grant number DMR 96-14999.

\bibliographystyle{prsty}

\begin{thebibliography}{10}

\bibitem{kondo_papers}
P.~W. Anderson, J. Phys. C {\bf 3},  2346  (1970);
 P.~W. Anderson, G. Yuval, and D.~R. Hamann, Phys. Rev. B {\bf 1},  4464
  (1970);
K.~G. Wilson, Rev. Mod. Phys. {\bf 47},  773  (1975);
N. Andrei, K. Furuya, and J.~H. Lowenstein, Rev. Mod. Phys. {\bf 55},  331
  (1983);
A.~M. Tsvelik and P.~B. Wiegmann, Adv. Phys. {\bf 32},  453  (1983);
I. Affleck, Acta Phys. Polon. B {\bf 26},  1869  (1995).

\bibitem{nozieres_multichannel}
P. Nozieres and A. Blandin, J. Phys. (Paris) {\bf 41},  193  (1980).

\bibitem{maple_nfl_review}
M.~B. Maple, J. Low Temp. Phys {\bf 99},  223  (1995).

\bibitem{cox_kondo_review}
D.~L. Cox and A. Zawadowski, Ann. Phys. {\bf 47},  599  (1998).

\bibitem{schulz_houches_revue}
H.~J. Schulz,  in {\em Mesoscopic quantum physics, Les Houches LXI}, edited by
  E. Akkermans, G. Montambaux, J.~L. Pichard, and J. Zinn-Justin (Elsevier,
  Amsterdam, 1995).

\bibitem{white}
S.~R. White, Phys. Rev. Lett. {\bf 69},  2863  (1992).

\bibitem{tsunetsugu_kondo_1d}
H. Tsunetsugu, M. Sigrist, and K. Ueda, Rev. Mod. Phys. {\bf 69},  809  (1997); N. Shibata and K. Ueda, J. Phys. Condens. Matter {\bf 11},  R1  (1999).

\bibitem{sikkema_spingap_kondo_1d}
A.~E. Sikkema, I. Affleck, and S.~R. White, Phys. Rev. Lett. {\bf 79},  929
  (1997).

\bibitem{exotic_pairing}
P. Coleman, A. Georges, and A. Tsvelik, J. Phys. Condens. Matter {\bf 79},  345
   (1997) ; O. Zachar and A.~M. Tsvelik, cond-mat/9909296
(unpublished).


\bibitem{witten_wz}
E. Witten, Commun. Math. Phys. {\bf 92},  455  (1984).

\bibitem{affleck_tsvelik}
I. Affleck,  in {\em Fields, Strings and Critical Phenomena}, edited by E.
  Brezin and J. Zinn-Justin (Elsevier Science Publishers, Amsterdam, 1988);
A. Tsvelik, {\em Quantum Field Theory in Condensed Matter Physics} (Cambridge
  University Press, Cambridge, 1995).


\bibitem{andrei_chiral_nfl}
N. Andrei, M.~R. Douglas, and A. Jerez, Phys. Rev. B {\bf 58},  7619  (1998).

\bibitem{goddard_unitar_repres_viras}
P. Goddard, A. Kent, and D. Olive, Commun. Math. Phys. {\bf 103},  105  (1986).

\bibitem{realiz} An interesting realization of the same
fixed point Hamiltonian in a frustrated three-chain ladder was recently 
given by P. Azaria, P. Lecheminant, 
and A.~A. Nersesyan, Phys. Rev. B {\bf 58},  R8881 (1998). The authors
also noticed the relation of their fixed point with the spin sector of
the two channel Kondo Heisenberg lattice.

\bibitem{kondo_2channel}
V.~J. Emery and S.~A. Kivelson, Phys. Rev. B {\bf 46},  10812  (1992);
A.~M. Sengupta and A. Georges, Phys. Rev. B {\bf 49},  10020  (1994);
P. Coleman, L.~B. Ioffe, and A.~M. Tsvelik, Phys. Rev. B {\bf 52},  6611
  (1995);
A.~J. Schofield, Phys. Rev. B {\bf 55},  5627  (1997).



\bibitem{odd_freq_pairing}
V.~L. Berezinskii, JETP Lett. {\bf 20},  287  (1974);
A. Balatsky and E. Abrahams, Phys. Rev. B {\bf 45},  13125  (1992).



\bibitem{yamanaka_luttinger_thm}
M. Yamanaka, M. Oshikawa, and I. Affleck, Phys. Rev. Lett. {\bf 79},  1110
  (1997).

\bibitem{zamolodchikov_c_thm}
A.~B. Zamolodchikov, JETP Lett. {\bf 43},  730  (1986).

\bibitem{orignac_andrei_unpublished}
E. Orignac and N. Andrei, 2000, article in preparation.

\bibitem{tsvelik_mck_comm}
A.~M. Tsvelik, Phys. Rev. Lett. {\bf 72}, 1048 (1994).

\bibitem{nozieres_exhaustion}
P. Nozi{\`e}res, Ann. Phys. (Paris) {\bf 10},  19  (1985);
P. Nozi{\`e}res, Eur. Phys. J. B {\bf 6},  447  (1998).

\end{thebibliography}

\begin{table}
\caption{The conformal weights of the operators  in the theory described by
(\ref{eq:chiral_theory}). Here, $\delta_N=\frac{3}{(N+1)(N+2)}$.
  For $N=2$, the conformal weights  of $\vec{\sigma}_L$ and
$\vec{\sigma}_R$ are 
respectively  $(0,1)$ and $(1,0)$.}
\begin{tabular}{cl}
\hline
Operator & Conformal weights\\
         & at the fixed point \\
\hline
$\psi_{R,n,\sigma}$ & $ \left(\frac 1 2 + \frac {\delta_N} 4 ,
  \frac {\delta_N} 4 \right)$ \\
$\psi_{L,n,\sigma}$ & $ \left( \frac {\delta_N} 4 ,
 \frac 1 2 + \frac {\delta_N} 4 \right)$ \\
$\tilde{g}_{L,\beta}$ & $ \left( \frac 3 {4(N+1)}, \frac 1 4 +\frac 3 {4(N+1)}  \right)$ \\
$\tilde{g}_{R,\beta}$ & $ \left( \frac 1 4 +\frac 3 {4(N+1)}, \frac 3 {4(N+1)}    \right)$ \\
$ \vec{\sigma}_L(x) $ & $\left(\frac 2 {N+1} ,1+\frac 2 {N+1} \right),
 N\ge 3$\\
$ \vec{\sigma}_R(x) $ & $\left(1+\frac 2 {N+1} ,\frac 2 {N+1} \right),
 N\ge 3$\\
$ \vec{S}_R(x)$ & $ \left(1,0\right)$ \\
$ \vec{S}_L(x)$ & $ \left(0,1\right)$ \\
$O_{\text{CDW}}=\psi^\dagger_{L,n,\sigma} \psi_{R,n,\sigma}$ &
 $\left(\frac{ 1 +\delta_N} 2 , \frac {1+\delta_N} 2 \right) $ \\
$ \vec{O}_{\text{SDW}}=\psi^\dagger_{L,n,\sigma}\vec{\sigma}_{\sigma,\sigma'}
 \psi_{R,n,\sigma'}$ & $\left(\frac{ 1 +\delta_N} 2 , \frac {1+\delta_N} 2 \right)$\\
$O_{\text{SS}}=-\imath \psi_{L,n,\sigma}\sigma^y_{\sigma,\sigma'}
\psi_{R,n,\sigma'} $ &$\left(\frac{ 1 +\delta_N} 2 , \frac {1+\delta_N} 2 \right)$ \\
$ \vec{O}_{\text{TS}}=-\imath 
\psi_{L,n,\sigma}(\vec{\sigma}\sigma_y)_{\sigma,\sigma'}
\psi_{R,n,\sigma'}$ &$\left(\frac{ 1 +\delta_N} 2 , \frac {1+\delta_N} 2 \right)$\\
$O_{\text{c-SP}}=\vec{n}.\vec{O}_{TS}$ & $\left( \frac 3 4 -
\frac{3}{2(N+2)}, \frac 3 4 -
\frac{3}{2(N+2)}\right)$  \\ 
$ O_{\text{c-CDW}}=\vec{n}.\vec{O}_{\text{SDW}}$ & $\left( \frac 3 4 -
\frac{3}{2(N+2)}, \frac 3 4 -
\frac{3}{2(N+2)}\right)$ \\
\hline
\end{tabular}
\label{tab:conformal_weights}
\end{table}

\end{document}